\documentclass[aps,prl,twocolumn,showpacs,preprintnumbers,amsmath,amssymb]{revtex4}

\usepackage{undertilde}
\newcommand\D{{\cal D}}
\newcommand\be{\begin{eqnarray}}
\newcommand\ee{\end{eqnarray}}

\begin{document}


\title{On deformations of Ashtekar's constraint algebra}

\author{Kirill Krasnov}
  \affiliation{School of Mathematical Sciences, University of Nottingham, Nottingham, NG7 2RD, UK}

\date{November 1, 2007}

\begin{abstract} We show that the constraint algebra of Ashtekar's Hamiltonian
formulation of general relativity can be non-trivially deformed by allowing the
cosmological constant to become an arbitrary function of the (Weyl) curvature.
Our result implies that there is not one but infinitely many (parameterized by
an arbitrary function) four-dimensional gravity theories propagating two
degrees of freedom.
\end{abstract}


\maketitle

In \cite{Ashtekar:1987gu} Ashtekar described a new Hamiltonian formulation of
general relativity (GR) in which the canonically conjugate phase-space
variables are a densitized triad $\tilde{\sigma}^{ai}$ and a (complexified)
${\rm SU}(2)$ connection $A_{a}^i$. Here spatial and ``internal'' indices are
denoted by lower-case Latin letters from the beginning and from the middle of
the alphabet, respectively. The constraints of general relativity take an
amazingly simple form in this formulation: \be\label{gauss} D_a
\tilde{\sigma}^{ai} &\approx& 0,
\\ \label{diffeo} \tilde{\sigma}^{ai} F_{ab}^i &\approx& 0, \\ \label{hamilt}
\epsilon^{ijk} \tilde{\sigma}^{ai} \tilde{\sigma}^{bj} F_{ab}^k + \Lambda
\epsilon^{ijk} \utilde{\epsilon}_{abc} \tilde{\sigma}^{ai} \tilde{\sigma}^{bj}
\tilde{\sigma}^{ck} &\approx& 0. \ee Here $D_a$ is the covariant derivative
with respect to the connection $A_{a}^i$, $\epsilon^{ijk}$ and
$\utilde{\epsilon}_{abc}$ are the completely anti-symmetric ``internal'' and
spatial tensors taking values $\pm 1$ (undertilde denotes the density weight
$-1$), and $F_{ab}^i=2\partial_{[a}A_{b]}^i + \epsilon^{ijk}A_a^j A_b^k$ is the
curvature of $A_a^i$. In the original work \cite{Ashtekar:1987gu} the
Hamiltonian formulation of GR with zero cosmological constant was given. A
generalization to the case of non-zero $\Lambda$ is trivial and consists in
adding the second term in (\ref{hamilt}). The constraints (\ref{gauss}),
(\ref{diffeo}) and (\ref{hamilt}) are the Gauss, diffeomorphism and Hamiltonian
ones, respectively. They form a first-class algebra, i.e., the Poisson bracket
of any two constraints vanishes on the constraint surface. This implies that
(\ref{gauss}), (\ref{diffeo}) and (\ref{hamilt}) generate gauge transformations
of the theory and allows a simple count of the number of physical (propagating)
degrees of freedom (DOF)\@. Specifically, one has $3\times 3=9$ kinematical
configurational variables $A_a^i$, together with three Gauss, three
diffeomorphism and one Hamiltonian constraint, which gives two physical DOF\@.

The main aim of this letter is to point out that there exists an
infinite-parametric family of deformations of the above constraint algebra. In
the deformed case the canonically conjugate variables are still the same
densitiezed triad $\tilde{\sigma}^{ai}$ and the connection $A_{a}^i$. The Gauss
and diffeomorphism constraints are unchanged. The only modification occurs in
the Hamiltonian constraint, in which the cosmological constant $\Lambda$ gets
replaced by an (arbitrary) function $\phi(\Psi)$ of the (symmetric) tensor:
\be\nonumber \Psi^{ij}:= \frac{(F^{(i}_{ab} \epsilon^{j)kl} \tilde{\sigma}^{ak}
\tilde{\sigma}^{bl})_{{\rm tr-free}}}
{\utilde{\epsilon}_{abc} \tilde{\sigma}^{ai} \tilde{\sigma}^{bj} \tilde{\sigma}^{ck}} = \\
\label{psi} (F^{(i}_{ab} \epsilon^{j)kl} \sigma^{ak} \sigma^{bl})_{{\rm
tr-free}}, \ee where ``tr-free'' denotes the trace-free part. Thus, the
modified Hamiltonian constraint becomes:
\begin{equation}\tag{\ref{hamilt}$'$}\label{hamilt-new}
\epsilon^{ijk} \tilde{\sigma}^{ai} \tilde{\sigma}^{bj} F_{ab}^k + \phi(\Psi) \epsilon^{ijk}
\utilde{\epsilon}_{abc} \tilde{\sigma}^{ai} \tilde{\sigma}^{bj} \tilde{\sigma}^{ck}
\approx 0.
\end{equation}
As we shall demonstrate, the algebra of the constraints (\ref{gauss}),
(\ref{diffeo}) and (\ref{hamilt-new}) is still of the first class, which
implies that the count of the number of physical degrees of freedom is
unchanged and the theory still propagates two DOF\@. In GR the quantity
$\Psi^{ij}$ defined by (\ref{psi}) is nothing but the Weyl part of the Riemann
curvature tensor. The modified theory (\ref{hamilt-new}) can thus be described
as one in which the cosmological constant became a non-trivial function of the
``curvature''. We shall also give the Lagrangian generally covariant
description of this class of modified gravity theories. Taken together, the
results reported in this letter imply that there is not one, but an infinite
number of four-dimensional generally covariant gravity theories propagating two
degrees of freedom. Let us also note that since there are only two independent
invariants ${\rm Tr}(\Psi)^2$, ${\rm Tr}(\Psi)^3$ that can be constructed from
a traceless symmetric matrix $\Psi$, the theories under consideration are
parameterized by a function $\phi(\Psi)$ of two variables.

\bigskip

Since only the Hamiltonian constraint has been modified we only have to verify
its algebra. However, because the Gauss and diffeomorphism constraints generate
gauge and spatial diffeomorphism transformations respectively, it is clear that their algebra with the new
Hamiltonian will not get modified. More precisely, as the Hamiltonian is gauge
invariant it commutes with the Gauss constraint, and the Poisson bracket of the
diffeomorphism and Hamiltonian constraints gives back the Hamiltonian with a
Lie derivative of the lapse function; see \cite{Ashtekar:1987gu} for details.
Thus, as in the case of GR, the only non-trivial task is to verify the algebra
of the Hamiltonian constraint with itself. For this purpose we, as usual, form
the generator of transformations by smearing the constraint (\ref{hamilt-new})
against a lapse function: \be {\cal C}_{\utilde{N}} := \int d^3x \,
\utilde{N}\, {\rm (Hamiltonian)}. \ee To ensure that the above integral
converges we assume the same fall-off conditions for the phase space variables
as in GR; see \cite{Ashtekar:1987gu}. To compute the Poisson bracket $\{ {\cal
C}_{\utilde{N}_1},{\cal C}_{\utilde{N}_2}\}$ we will need the variational
derivatives of the smeared constraint. These are obtained from the following
useful formula: \be \nonumber \delta {\cal C}_{\utilde{N}} = \int d^3x
\, \utilde{N}\, \Big( h_{ij} \delta(F^{(i}_{ab} \epsilon^{j)kl}
\tilde{\sigma}^{ak} \tilde{\sigma}^{bl})+
\\ \label{variation} 
(\phi(\Psi) - M_{lm}\Psi^{lm})
\delta(\utilde{\epsilon}_{abc} \epsilon^{ijk} \tilde{\sigma}^{ai} \tilde{\sigma}^{bj} \tilde{\sigma}^{ck})\Big).
\ee
Here we have introduced:
\be\label{M}
M_{ij} := \frac{\partial\phi(\Psi)}{\partial \Psi^{ij}}, \\ \label{h}
h_{ij}:= \delta_{ij} + M_{ij}.
\ee
Note that the matrix $M_{ij}$ is symmetric traceless. For this reason we can
replace $M_{ij}\delta(\ldots)^{ij}_{\rm tr-free}$ by the full variation
 $M_{ij}\delta(\ldots)^{ij}$, and this is what was done to get (\ref{variation}).
This gives the following expressions for the variational derivatives:
\be\label{var-a}
\frac{\delta {\cal C}_{\utilde{N}}}{\delta A_a^i} = 2\D_b\left( \utilde{N}
h_{ij} \epsilon^{jkl} \tilde{\sigma}^{ak} \tilde{\sigma}^{bl}\right), \\
\label{var-s} \frac{\delta {\cal C}_{\utilde{N}}}{\delta \tilde{\sigma}^{ai}}=
2 \utilde{N} h_{jl} F^j_{ab} \epsilon^{ikl} \tilde{\sigma}^{kb} + \nonumber \\
3 \utilde{N} (\phi(\Psi) - M_{lm}\Psi^{lm}) \utilde{\epsilon}_{abc}
\epsilon^{ijk} \tilde{\sigma}^{bj} \tilde{\sigma}^{ck}.
\ee
When we substitute these into the formula
\be
\{ {\cal C}_{\utilde{N}_1},{\cal C}_{\utilde{N}_2}\} = \int d^3x \left(
\frac{\delta {\cal C}_{\utilde{N}_1}}{\delta A_a^i} \frac{\delta {\cal
C}_{\utilde{N}_2}}{\delta \tilde{\sigma}^{ai}}- \frac{\delta {\cal
C}_{\utilde{N}_2}}{\delta A_a^i} \frac{\delta {\cal C}_{\utilde{N}_1}}{\delta
\tilde{\sigma}^{ai}}\right),
\ee
the second term in (\ref{var-s}) does not contribute. This happens for exactly the same
reason as in GR: the symmetric tensor $h_{ij}$ gets contracted with the
anti-symmetric tensor $\epsilon^{ijk}$. Thus, we only have to worry
about the first term in (\ref{var-s}). The Poisson bracket is then
equal to four times the integral over the spatial manifold of the following quantity:
\be\nonumber
\D_b\left( \utilde{N}_1 h_{ij} \epsilon^{jkl}  
\tilde{\sigma}^{ak} \tilde{\sigma}^{bl}\right) 
\utilde{N}_2 h_{mp} F^m_{ac} \epsilon^{inp} \tilde{\sigma}^{cn} - (1\leftrightarrow 2)=\\
\nonumber
\utilde{\utilde{N}}_b h_{ij} \epsilon^{jkl}  \tilde{\sigma}^{ak} \tilde{\sigma}^{bl}
h_{mp} F^m_{ac} \epsilon^{inp} \tilde{\sigma}^{cn},
\ee
where
\be
\utilde{\utilde{N}}_a =\partial_a(\utilde{N}_1) \utilde{N}_2  - 
  \partial_a(\utilde{N}_2) \utilde{N}_1.
\ee
In order to simplify this, let us replace
$F^m_{ac} \tilde{\sigma}^{ak} \tilde{\sigma}^{cn}$ by $(1/2) \epsilon^{qkn} (F\sigma\sigma)^{mq}$,
where 
\be\label{fss}
(F\sigma\sigma)^{ij}:=F^{i}_{ab} \epsilon^{jkl} \tilde{\sigma}^{ak} \tilde{\sigma}^{bl}.
\ee
Expanding $\epsilon^{jkl}\epsilon^{qkn}$ and using the fact that $h_{ij}$ is a symmetric matrix,
after some simple algebra we can transform the Poisson bracket of interest to the following
simple form:
\be\nonumber
2\tilde{\sigma}^{bl}\epsilon^{lij}  h_{jn} h_{im} \utilde{\utilde{N}}_b (F\sigma\sigma)^{mn}.
\ee
It is clear that only the anti-symmetric part of $(F\sigma\sigma)$ contributes and that
this expression is proportional to the diffeomorphism constraint since
$(1/2) \epsilon^{ijk} (F\sigma\sigma)^{jk}=\tilde{\sigma}^{ai}\tilde{\sigma}^{bj}F^j_{ab}$.
This allows us to write down our final result for the commutator of two Hamiltonian constraints:
\be
\{ {\cal C}_{\utilde{N}_1},{\cal C}_{\utilde{N}_2}\} = 4 \int d^3x\, \tilde{\tilde{Q}}^{ab}
\utilde{\utilde{N}}_b F^i_{ac} \tilde{\sigma}^{ci},
\ee 
where
\be
\tilde{\tilde{Q}}^{ab} := \frac{1}{2} \tilde{\sigma}^{ai} \tilde{\sigma}^{bl} \epsilon^{ijk} \epsilon^{lmn} h_{jm} h_{kn}.
\ee
The expression for the Poisson bracket is thus the same as in the GR case, see
\cite{Ashtekar:1987gu}, apart from the fact that a more complicated metric
$\tilde{\tilde{Q}}^{ab}$ is used in place of $\tilde{\sigma}^{ai}
\tilde{\sigma}^{bi}$. When $\phi=\mbox{const}$, the matrix $h_{ij}=\delta_{ij}$
and one gets back the familiar GR result. This finishes our demonstration of
the fact that the algebra of the constraints (\ref{gauss}), (\ref{diffeo}) and
(\ref{hamilt-new}) is of the first class.

\bigskip

It is well known (see, e.g., \cite{CDJ}) that Ashtekar's constraint algebra
arises as a result of the 3+1 decomposition of the Pleba\'nski formulation
\cite{Plebanski} of GR\@. Here we describe the spacetime covariant theory that
leads to the modified constraint algebra (\ref{gauss}), (\ref{diffeo}) and
(\ref{hamilt-new}). This theory was proposed in \cite{Krasnov:2006du} and its
action is given by:
\be\label{action}
S= \int B_i \wedge F^i(A) - \frac{1}{2}
\left( \Psi_{ij} - \frac{1}{3} \delta_{ij} \phi(\Psi)\right) B^i\wedge B^j.
\ee
When $\phi= \mbox{const} =\Lambda$ one recognizes in this the Pleba\'nski
action for GR with the cosmological constant. For $\phi$ depending
non-trivially on the Lagrange multiplier field $\Psi^{ij}$ one gets a new
theory. In the above action $B^i$ is an ${\rm SU}(2)$ Lie algebra valued
two-form, $A^i$ is the connection one-form, $F^i(A)$ is its curvature, and
$\Psi^{ij}$ is the Lagrange multiplier field, which is required to be symmetric
traceless.

The 3+1 decomposition of the theory (\ref{action}) proceeds as in \cite{CDJ}.
One easily finds that the momentum conjugate to the spatial connection $A_a^i$
is given by $\tilde{\sigma}^{ai}=\tilde{\epsilon}^{abc}B_{bc}^i$ and that the
term in the action containing $A_0^i$ gives the standard expression for the
Gauss constraint. Using the momentum variable, one re-writes the expression for the
Lagrange multiplier term as:
\be\label{1}
-\left( \Psi_{ij} - \frac{1}{3} \delta_{ij} \phi(\Psi)\right)
B_{0a}^i \tilde{\sigma}^{aj}.
\ee
The other term we have to consider is
\be\label{2}
2B_{0a}^i \tilde{\epsilon}^{abc} F_{bc}^i.
\ee
The quantity $B_{0a}^i$ plays the role of a Lagrange multiplier and 
is a spatial one-form with values in the Lie algebra. One
can use the triad to convert it to a quantity $B_0^{ij}$ with two internal indices.
This can be decomposed into its symmetric trace and traceless as well as anti-symmetric parts.
We get the following very useful representation:
\be\label{b0}
B_{0a}^i = (\delta^{ij}+M^{ij})\utilde{\sigma}_a^j \utilde{N} \tilde{\tilde{\sigma}}
+ \epsilon^{ijk} \utilde{\sigma}_a^j \tilde{N}^k,
\ee
where we have introduced the co-triad $\utilde{\sigma}_a^i$ of density weight $-1$ satisfying
$\tilde{\sigma}^{ai} \utilde{\sigma}_a^j = \delta^{ij}$, densitiezed lapse $\utilde{N}$ and
shift $\tilde{N}^k$ functions, as well as the determinant of the spatial metric
$\tilde{\tilde{\sigma}}=\utilde{\epsilon}_{abc} \tilde{\sigma}^{ai} \tilde{\sigma}^{bj} \tilde{\sigma}^{ck}$.
The matrix $M^{ij}$ in (\ref{b0}) is traceless and is (proportional to) the traceless part
of $B_0^{(ij)}$. Using this representation it is easy to compute the Lagrange
multiplier term (\ref{1}). We get:
\be\label{2'}
(\phi(\Psi) - M_{ij} \Psi^{ij}) \utilde{N} \tilde{\tilde{\sigma}}.
\ee
Varying the corresponding term in the action with respect to $\Psi$ we get the
equation that states that $M_{ij}$ is precisely the matrix introduced in
(\ref{M}) above.

Let us now consider an equation one gets by varying the action with respect to
the fields $M^{ij}$. There is a contribution from the term (\ref{2}) as well as
from (\ref{2'}). It is not hard to see that the resulting equation is precisely
(\ref{psi}) relating the Lagrange multiplier field to other phase space
variables. A moment of reflection shows that the system of constraints
(\ref{psi}), (\ref{M}) as well as $P_{\Psi}\approx 0$, $P_{M}\approx 0$, where
$P_{\Psi}$ and $P_{M}$ are the momenta conjugate to $\Psi$ and $M$,
respectively, is of second class. Thus, these variables should be eliminated by
solving for them and substituting the result into the action. When doing this
we find that the terms involving $M^{ij}$ cancel, and the action in the
Hamiltonian form is:
\be\nonumber
S=\int d^3x\, \Big( \tilde{\sigma}^{ai} \dot{A}_a^i + A_0^i \D_a \tilde{\sigma}^{ai} +
\\ 
(\delta^{ij} \utilde{N} + \frac{1}{2} \epsilon^{ijk} \utilde{N}^k) (F\sigma\sigma)^{ij} +
\utilde{N} \phi(\Psi) \tilde{\tilde{\sigma}} \Big),
\ee
where the matrix $(F\sigma\sigma)^{ij}$ was defined in (\ref{fss}). We thus get
precisely the structure of the phase space as described above.

\bigskip

We have described an infinitely large family of four-dimensional modified
gravity theories propagating two degrees of freedom. A theory from the family
is parameterized by an arbitrary function $\phi(\Psi)$ of a symmetric traceless
matrix (thus a function of two variables). General relativity itself belongs to
this family as the simplest case $\phi= \mbox{const} =\Lambda$. The essence of
modification is to replace the cosmological constant by a ``cosmological
function'' of the curvature. The results presented here are particularly
surprising in view of the well-known fact that the only four-dimensional
generally covariant theory {\it of spacetime metric} propagating two degrees of
freedom is general relativity. The theories described here avoid this no-go
theorem because the spacetime metric emerges in them only as a derived concept.

The described new class of gravity theories may have far-reaching applications.
In particular, it may be possible to use them to address the
cosmological-constant problem, which is to explain why the observed
cosmological constant is so different from the one that is expected to arise as the vacuum energy of
quantum fields. One possibility that arises in the present context is that the
vacuum energy of fields with typical energy scale, say, $1/l_p^2$, where $l_p$
is the Planck length, contributes the natural value $1/l_p^2$ to
$\phi(1/l_p^2)$ for curvatures $\Psi$ of the order of $1/l_p^2$. This large
value, however, does not have to be equal to $\phi(0)$, which plays the role in
cosmology. Some other possible physical applications of this class of theories
have been described in \cite{Krasnov:2007ky}.

Our result also has implications for the program of loop quantum gravity, see e.g. \cite{LQG}.
This program is often criticized for elevating general relativity to the fundamental
status even at the Planck scale, where, as is widely believed, the GR description is invalid.
Our result shows that the starting point of loop quantum gravity, i.e., the kinematical
phase space together with the constraint algebra of the type described above, is common to a
much more general class of theories than GR. Thus, provided one allows for a more
general class of Hamiltonian constraints, the technology and all the results of
loop quantum gravity immediately extend to quite a large class of theories, some
of them being possibly much closer to the correct description of gravity at Planck scale.

Our final remark is that some of the results of this letter also follow from the analysis
of \cite{Bengtsson:2007zx} performed in the ``pure connection'' formulation.
However, the description of the constraint algebra given here is new. Our description also emphasizes the
fact that the only change as compared to GR is in allowing the cosmological constant
to become the ``cosmological function'', something that is not at all clear from
\cite{Bengtsson:2007zx}.

The author would like to thank Yuri Shtanov for reading the manuscript. This work is
supported by an EPSRC Advanced Fellowship.

\end{document}